\documentclass[12pt]{article}
\usepackage{graphicx}
\usepackage{xcolor}
\begin{document}
\title{
Dynamical Mass Growing of Fermion with Bare Mass 
in Two Dimensions
}
\author{
Toyoki Matsuyama \thanks{e-mail: matsuyat@nara-edu.ac.jp} \\
Department of Physics, Nara University of Education \\
Takabatake-cho, Nara 630-8528, Japan \\
and \\
Center of Education and Research for \\
Topological Science and Technology, \\
Hokkaido University, Sapporo, \\
Hokkaido 060-8890, Japan \\
}
\baselineskip=21.5pt
\date{\today}
\maketitle
\begin{abstract}
We study dynamical mass generation of a fermion with and without a bare mass 
by coupling with a massive vector field in two-dimensional space-time.  
To estimate a non-perturbative effect on the fermion mass, we employ the 
Schwinger-Dyson equations in the lowest-ladder approximation, which are solved 
by an approximated analytical method and also by a numerical method.  
We define a $purely$ dynamical mass as a remnant after subtracting the bare 
mass from a total dynamical mass.  
We clarify dependence of the purely dynamical mass on the bare mass of the 
fermion in various region of a coupling constant.  
Especially we find that the purely dynamical masses growing from the different 
bare masses coincide with each other at a specific value of the coupling 
constant where a kind of a $duality$ relation on the bare masses is  
satisfied.   
\end{abstract}

\section{Introduction}

A mass is one of the most important physical quantities in particle 
physics.  
To explain an origin of mass, a plausible scenario has been considered in 
quantum field theories.  
It is a dynamical mass generation:  
Without interactions, a fundamental particle has no mass because 
the mass is forbidden by a symmetry.  
Even if a radiative correction is considered perturbatively, the mass takes 
still zero value as far as the interaction itself does not violate the 
symmetry.  
It is a non-perturbative effect that creates the mass of the particle.  
This scenario is very fascinating for us and many works in various models 
have been done until now.  
In these works, we have found many interesting aspects [3], [4], [5], [6] on the dynamical mass 
generation.

What we consider in this paper is based on a different standing point.  
We wonder what is happened in the dynamical mass generation, if the particle 
has a $bare$ mass [7], [8], [9]. 
The introduction of the bare mass might be consider to be unfavorable for  
the philosophy of the dynamical mass generation.  
But we think that it is still meaningful to consider the bare mass from  
the following points of view, at least.  

Until now, four types of fundamental interactions in Nature are known.  
These interactions may be unified at higher energy scales.  
Among these fundamental interactions at some unified scales, there may 
exist a specific one which is most responsible for the dynamical mass 
generation.  
It might be the gravitational interaction or a more fundamental interaction 
in which the gravity is unified.  
The interaction may create the mass from the null.  
The mass so created may be considered as a "seed" or "core" and will be 
corrected radiatively by other interactions.  
Then the particle will get an observed amount of the mass.  
These situation motivates us to study the dynamical mass generated from a 
bare mass.

From a practical point of view, it may be still interesting to study how the 
amount of the bare mass affects the dynamical mass generation.  
In studies of the standard model, we need to assign quarks intrinsic bare  
masses to explain, e.g., the masses of hadrons.  
In a condensed matter system, we consider that an electron is massive and 
study an effective mass of the electron in a matter.  
For these investigations, it is valuable to understand a mechanism of the 
dynamical growing of the fermion mass with bare mass.  

In this paper, we consider a two-dimensional quantum electrodynamics where the 
massive fermion is coupled to the massive vector field.  
This model is rather simple and useful as a laboratory to understand an 
essential of the mechanism.   
It also should be noticed that the model itself has some important characters 
as, e.g., confinement or chiral symmetry breaking.  
The study of the model will be useful for understanding a quantum 
chromodynamics [10], [11], [12].
(An extension to various model in other dimensions will be reported 
elsewhere.)  
In addition, the model may have an important application to the condensed 
matter systems where pseudo-one-dimensional systems are realized in specific 
matters and interesting phenomena like the Haldane gap or the 
Tomonaga-Luttinger liquid are extensively studied.   

To study non-perturbative effects on the fermion mass, we use the 
Schwinger-Dyson method [13].
We derive the Schwinger-Dyson equation in the quenched ladder approximation, 
in which a vacuum polarization effect is neglected.  
The equation is consistent with the Ward-Takahashi identity in the Feynman 
gauge so that we proceed our analysis in this gauge.  
(The Schwinger-Dyson equation in a general gauge is presented in Appendix A.) 
Of course, it must be better to consider the vacuum polarization effect but, 
for the massive fermion, it is very difficult technically as following 
reasons: 
An angular integration in the Schwinger-Dyson equation cannot be 
accomplished analytically because the vacuum polarization effect due to the 
massive fermion has a complicated form.  
If we take a massless limit, then the integrations can be done completely.  
In the massive case which we are interested in, there remain double integrals 
in the Schwinger-Dyson equation so that the analysis becomes very hard.  
Therefore we limit ourselves to the quenched approximation in this paper. 

This paper is organized as follows.  
In Sec. 2, we define the model which we study.  
In Sec. 3, the Schwinger-Dyson equation is derived in the lowest ladder 
approximation.  
In Sec. 4, the equation is analytically solved in a constant approximation.  
A total mass and a purely dynamical mass which will be defined newly there are 
studied.  
In Sec. 5, a numerical study is presented and compared with the analytical 
one.  
Conclusions with discussions are given in Sec.6.       
Some details of calculations and useful formulae are given in Appendices.  

\section{Massive Proca $QED_2$}

In this paper, we consider a massive Proca quantum electrodynamics in 
(1+1)-dimensional space-time (massive Proca $QED_2$) [13].
The model is defined by a Lagrangian density as 
%%%%%%%%%%%%%%%%%%%%%%%%%%%%%%%%%%%%%%%%%%%%%%%%%%%%%%%%%%%%%%%%%%%%%%%%%%%%%
\begin{eqnarray}
{\cal L} = - \frac{1}{4} F_{\mu\nu} F^{\mu\nu} 
           + \frac{1}{2} \mu^{2} A_{\mu} A^{\mu} 
           - \frac{1}{2 \lambda} (\partial_{\mu}A^\mu)^2 
           + \bar{\psi} (i \not \! \partial - e \not \! \! A-m) \psi \ \ . 
\label{lag}
\end{eqnarray}
%%%%%%%%%%%%%%%%%%%%%%%%%%%%%%%%%%%%%%%%%%%%%%%%%%%%%%%%%%%%%%%%%%%%%%%%%%%%%
$e$ is a coupling constant with the dimensions of mass and $\lambda$ is a 
gauge-fixing parameter. 
$\psi$ is a Dirac field which has a bare mass $m$.  
In (1+1)-dimensions, Dirac matrices and their algebra are given by 
%%%%%%%%%%%%%%%%%%%%%%%%%%%%%%%%%%%%%%%%%%%%%%%%%%%%%%%%%%%%%%%%%%%%%%%%%%%%%
\begin{eqnarray}
\gamma^{0} = \sigma_{2} \ , \ \ \gamma^{1} = i \sigma_{1} \ , \ \ 
\gamma^{5} = \sigma_{3} = \gamma^0 \gamma^1 \ , 
\nonumber \\ 
\{ \gamma^{\mu},\gamma^{\nu} \} = 2 g^{\mu \nu} \ , \ \ 
\{ \gamma^{\mu}, \gamma^{5} \} =0 \ , 
\nonumber \\ 
\gamma^{\mu} \gamma^{\nu} = \epsilon^{\mu \nu} \gamma^{5} + g^{\mu \nu} \ , \ \
\gamma^{\mu} \gamma^{5} = - \epsilon^{\mu \nu} \gamma_{\nu} \ , \ \
(\gamma^{5})^{2}=1 \ \ .
\nonumber
\end{eqnarray}
%%%%%%%%%%%%%%%%%%%%%%%%%%%%%%%%%%%%%%%%%%%%%%%%%%%%%%%%%%%%%%%%%%%%%%%%%%%%%
On the other hand, $A^{\mu}$ is a massive vector field, called the Proca field, with a  mass $\mu$ which is introduced to rescue the theory from the infrared 
catastrophe.     
 
The model has three parameters with the dimensions of mass as $m$, $\mu$, and 
$e$.  
If we take a strong coupling limit as $m/e \rightarrow 0$ (and 
$\mu/e \rightarrow 0$), the model reduces to the Schwinger model.  
In a weak coupling limit as $e/m \rightarrow 0$, the model becomes the free 
theory.  
Thus the model are exactly solvable in both limits.  
Our interest is in an intermediate region of $m/e$.  
    
\section{Schwinger-Dyson Equation}

In order to estimate the dynamical fermion mass induced by the 
non-perturbative effect, we use the Schwinger-Dyson method [1], [2].
The so-called Schwinger-Dyson equation for a self-energy of the fermion is 
given by 
%%%%%%%%%%%%%%%%%%%%%%%%%%%%%%%%%%%%%%%%%%%%%%%%%%%%%%%%%%%%%%%%%%%%%%%%%%%%%
\begin{eqnarray}
\Sigma(p)=-im+(-ie)^2 \int \frac{d^2k}{(2\pi)^2} \gamma^\mu i S_F(k) 
\Gamma^{\nu} (k, p-k) i D'_{\mu \nu}(p-k) \ \ ,
\label{fullSD}
\end{eqnarray}
%%%%%%%%%%%%%%%%%%%%%%%%%%%%%%%%%%%%%%%%%%%%%%%%%%%%%%%%%%%%%%%%%%%%%%%%%%%%%
where $iD'_{\mu \nu}$ and $S_F$ are the full propagators of the gauge and 
fermion fields respectively.  
$\Gamma^{\mu}$ is a full vertex function.
Since it is impossible to solve this equation, we need to introduce an 
approximation.  
In this paper, we limit ourselves to use the lowest-ladder approximation 
because of the technical difficulty as explained in Sec.1.  
We replace the full vertex function $\Gamma^\mu$ to the bare vertex 
$\gamma^{\mu}$ and the full propagator of the gauge field $iD'^{\mu \nu}$ 
to the free propagator given by 
%%%%%%%%%%%%%%%%%%%%%%%%%%%%%%%%%%%%%%%%%%%%%%%%%%%%%%%%%%%%%%%%%%%%%%%%%%%%%
\begin{eqnarray}
iD^{\mu \nu}(k) = - i \left( \frac{g^{\mu \nu} - k^{\mu} k^{\nu} / \mu^2}
                                  {k^2-\mu^2}
                + \frac{k^{\mu} k^{\nu} / \mu^2}{k^2 - \lambda \mu^2} 
                \right) \ \ ,
\label{D}
\end{eqnarray}
%%%%%%%%%%%%%%%%%%%%%%%%%%%%%%%%%%%%%%%%%%%%%%%%%%%%%%%%%%%%%%%%%%%%%%%%%%%%%
which is derived from the Lagrangian density (\ref{lag}).  
Thus the Schwinger-Dyson equation in the lowest-ladder approximation becomes 
%%%%%%%%%%%%%%%%%%%%%%%%%%%%%%%%%%%%%%%%%%%%%%%%%%%%%%%%%%%%%%%%%%%%%%%%%%%%%
\begin{eqnarray}
\Sigma(p)=-im+(-ie)^2 \int \frac{d^2k}{(2\pi)^2} \gamma^\mu i S_F(k) 
\gamma^\nu i D_{\mu \nu}(p-k) \ \ .
\label{SD}
\end{eqnarray}
%%%%%%%%%%%%%%%%%%%%%%%%%%%%%%%%%%%%%%%%%%%%%%%%%%%%%%%%%%%%%%%%%%%%%%%%%%%%%

The general form of the full fermion propagator, required by the Lorentz 
covariance, is written as 
%%%%%%%%%%%%%%%%%%%%%%%%%%%%%%%%%%%%%%%%%%%%%%%%%%%%%%%%%%%%%%%%%%%%%%%%%%%%%
\begin{eqnarray}
i S_F(p) = \frac{i}{A(p) \not \! p - B(p)} 
         = \frac{i}{\not \! p - i \Sigma(p)} \ \ , 
\label{SF}
\end{eqnarray}
%%%%%%%%%%%%%%%%%%%%%%%%%%%%%%%%%%%%%%%%%%%%%%%%%%%%%%%%%%%%%%%%%%%%%%%%%%%%%
where $A(p)$ and $B(p)$ are functions of $p^2=p^{\mu} p_{\mu}$ while 
$\Sigma(p)$ is the function of $p^{\mu}$'s.  
$A(p)^{-1}$ is the renormalization factor of the wave function.   
We define the dynamical mass $m_{phys}$ by $m_{phys}=B(0)/A(0)$ as usual.  

The Schwinger-Dyson equation (\ref{SD}) can be separated into two coupled 
integral equations for $A(p)$ and $B(p)$ by using relations as   
%%%%%%%%%%%%%%%%%%%%%%%%%%%%%%%%%%%%%%%%%%%%%%%%%%%%%%%%%%%%%%%%%%%%%%%%%%%%%
\begin{eqnarray}
tr(\Sigma) = -2 i B \ , \ \ 
tr(\not \! p \Sigma) = 2 i (A-1) p^2 \ \ .
\label{relation}
\end{eqnarray}
%%%%%%%%%%%%%%%%%%%%%%%%%%%%%%%%%%%%%%%%%%%%%%%%%%%%%%%%%%%%%%%%%%%%%%%%%%%%%
In deriving the integral equations, we change the Minkowski metric to the 
Euclidean one by the Wick rotation as
 $k_0 \rightarrow i k_0$ and $\ p_0 \rightarrow i p_0$.  
Then we transform the integral variables from Cartesian coordinates 
to the polar ones.   
The integrals over angular variables can be done explicitly.  
After that, we obtain  
%%%%%%%%%%%%%%%%%%%%%%%%%%%%%%%%%%%%%%%%%%%%%%%%%%%%%%%%%%%%%%%%%%%%%%%%%%%%%
\begin{eqnarray}
B(p) &=& m+\frac{e^2}{2\pi} \int^\infty_{0} dk \frac{kB(k)}{A(k)^2 k^2+B(k)^2} 
     \left[ \frac{1}{\sqrt{(p^2+k^2+\mu^2)^2-4p^{2}k^2}} \right. 
\nonumber \\
       & &  \left. + \lambda
                 \frac{1}{\sqrt{(p^2+k^2+ \lambda \mu^2)^2-4p^2 k^2}} 
     \right] \ \ ,
\label{B} \\
A(p) &=& 1 - \frac{e^2}{4\pi} \frac{1}{p^2} \int^\infty_{0} dk 
             \frac{kA(k)}{A(k)^2 k^2+B(k)^2} \frac{1}{\mu^2} 
            \left[ \frac{(p^2+k^2)^2+(p^2+k^2)\mu^2-4k^2 p^2} 
                     {\sqrt{(p^2+k^2+\mu^2)^2-4k^2 p^2}} \right. 
\nonumber \\
        & &  \left. - \frac{(p^2+k^2)^2+(p^2+k^2)\lambda \mu^2-4k^2 p^2}
             {\sqrt{(p^2+k^2+ \lambda \mu^2)^2-4k^{2}p^2}} \right] \ \ .
\label{A}
\end{eqnarray}
%%%%%%%%%%%%%%%%%%%%%%%%%%%%%%%%%%%%%%%%%%%%%%%%%%%%%%%%%%%%%%%%%%%%%%%%%%%%%
(See Appendix A for the details.)  

We are now using the lowest-ladder approximation in which the full vertex 
function is replaced by the bare one.  
To be consistent with the Ward-Takahashi identity, there should be no 
wavefunction renormalization.  
We notice that Eq. (\ref{A}) reduces to $A(p)=1$ in the Feynman gauge 
($\lambda=1$).  
Therefore we proceed our analysis in the Feynman gauge, where Eq. (\ref{B}) 
is written as 
%%%%%%%%%%%%%%%%%%%%%%%%%%%%%%%%%%%%%%%%%%%%%%%%%%%%%%%%%%%%
\begin{eqnarray}
B(p)=m+\frac{e^2}{2\pi} \int^\infty_{0} dk \frac{2kB(k)}{k^2+B(k)^2} 
            \frac{1}{ \sqrt{(p^2+k^2+\mu^2)^2-4p^2 k^2} } \ \ . 
\label{Fey}
\end{eqnarray}
%%%%%%%%%%%%%%%%%%%%%%%%%%%%%%%%%%%%%%%%%%%%%%%%%%%%%%%%%%%%

Since there exist three parameters $e$, $m$, and $\mu$ which have the 
dimensions of mass, the model is controlled by two dimensionless parameters.  
By changing $p$ and $k$ to the dimensionless variables $s$ and $t$ 
defined by $p^2=\mu^2 s$ and $k^2=\mu^2 t$, we can rewrite Eq. (\ref{Fey}) 
to a dimensionless form as  
%%%%%%%%%%%%%%%%%%%%%%%%%%%%%%%%%%%%%%%%%%%%%%%%%%%%%%%%%%%%%%%%%%%%%%%%%%%%%
\begin{eqnarray}
b(s)=M+ g \int^\infty_{0} dt \frac{b(t)}{t+b(t)^2} 
            \frac{1}{\sqrt{(s+t+1)^2-4st} } \ \ ,
\label{dimless}
\end{eqnarray}
%%%%%%%%%%%%%%%%%%%%%%%%%%%%%%%%%%%%%%%%%%%%%%%%%%%%%%%%%%%%%%%%%%%%%%%%%%%%%
where $b(s)$, $g$ and $M$ defined by 
%%%%%%%%%%%%%%%%%%%%%%%%%%%%%%%%%%%%%%%%%%%%%%%%%%%%%%%%%%%%
\begin{eqnarray}
b(s) \equiv \frac{B(\mu\sqrt{s})}{\mu} \ , \ \
g \equiv \frac{e^2}{2 \pi \mu^2} \ , \ \ 
M \equiv \frac{m}{\mu} \ ,
\label{dimlessqu}
\end{eqnarray} 
%%%%%%%%%%%%%%%%%%%%%%%%%%%%%%%%%%%%%%%%%%%%%%%%%%%%%%%%%%%%
are dimensionless.  

From Eq. (\ref{Fey}), we can derive the upper bound for $B(p)$ as 
%%%%%%%%%%%%%%%%%%%%%%%%%%%%%%%%%%%%%%%%%%%%%%%%%%%%%%%%%%%%
\begin{eqnarray}
B(p) \leq m+\frac{1}{4} \frac{e^2}{\mu} \ \ ,
\label{upperB}
\end{eqnarray}
%%%%%%%%%%%%%%%%%%%%%%%%%%%%%%%%%%%%%%%%%%%%%%%%%%%%%%%%%%%%
and for $b(0)$ as
%%%%%%%%%%%%%%%%%%%%%%%%%%%%%%%%%%%%%%%%%%%%%%%%%%%%%%%%%%%%
\begin{eqnarray}
b(p) \leq M + \frac{\pi}{2} g \ \ . 
\label{upperb}
\end{eqnarray} 
%%%%%%%%%%%%%%%%%%%%%%%%%%%%%%%%%%%%%%%%%%%%%%%%%%%%%%%%%%%%
(See Appendix B for a derivation.)  
The bound is decided by $m$, $e^2$, and $\mu$.  
Thus the dynamical mass generation is enhanced if $m$ or $e^2$ is large, 
but is suppressed if 
the bare mass of the vector field $\mu$ is large, in reflecting that a virtual 
emission of a vector field is reduced if it is heavy.    
  
In the followings, we solve Eq. (\ref{dimless}) by an approximated 
analytical method and also by a numerical method.  

\section{Approximated Analytical Study}

Before proceeding to a numerical evaluation, we make a rough estimation of 
$b(0)$ by an approximated analytical method.  
It might be crude but will be helpful as reference for the complicated 
numerical study.  

\subsection{Constant Approximation} 
First, we put $s=0$ in Eq. (\ref{dimless}).  
Because the kernel dumps rapidly as $t$ increases, the most dominant 
contribution in the integral comes from the region of $t \approx 0$ so that 
we approximate $b(t)$ by $b(0)$ in the integral.  
We call this a "constant approximation".  
Then the integral on $t$ can be done explicitly (see also Appendix C) and we obtain 
%%%%%%%%%%%%%%%%%%%%%%%%%%%%%%%%%%%%%%%%%%%%%%%%%%%%%%%%%%%%
\begin{eqnarray}
b(0) = M + g \frac{b(0)}{b(0)^2-1} \ln b(0)^2 \ \ . 
\label{b}
\end{eqnarray} 
%%%%%%%%%%%%%%%%%%%%%%%%%%%%%%%%%%%%%%%%%%%%%%%%%%%%%%%%%%%%
At a first glance, $b(0)=M$ is a solution of Eq. (\ref{b}) when $g=0$, 
where $M$ is the dimensionless bare mass.  
It is expected because there should be no extra mass generation when the 
interaction is turned off.  
In the following successive subsections, we will see that fruitful 
information on the dynamical mass generation can be derived from this rather 
simple equation (\ref{b}), which will be confirmed by the numerical 
calculation in Sec. 5.  

\subsection{Total Mass}

\begin{figure}[h]
     \centering
     \includegraphics[width=10cm]{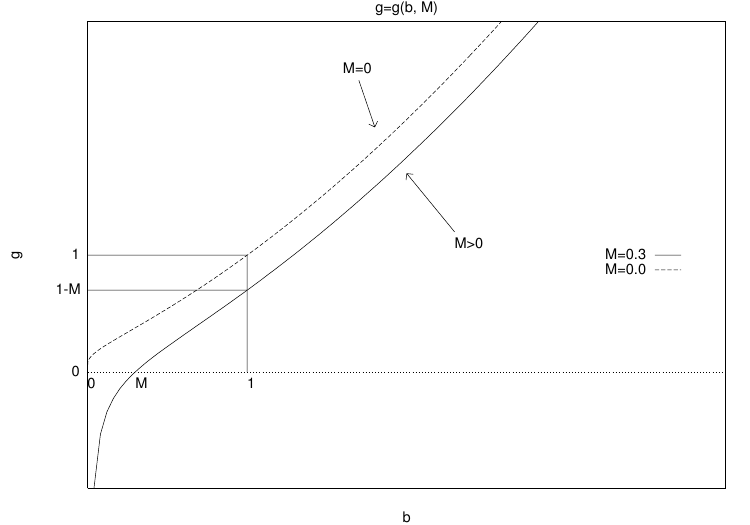}
\caption{ 
$g=g(b, M)$ for the cases of $M=0$ and $M = 0.3$, which have different 
asymptotic behaviour for $g <<1$.
}
\label{fig01}
\end{figure}

For a general $g$ ($\geq 0$ as defined by Eq. (\ref{dimlessqu}) ), we cannot 
express $b$ by using elementary functions of $g$.  
Instead, we take another strategy.  
We can solve Eq. (\ref{b}) for $g$ as a function of $b$ and $M$ explicitly.  
It is given by 
%%%%%%%%%%%%%%%%%%%%%%%%%%%%%%%%%%%%%%%%%%%%%%%%%%%%%%%%%%%%%%%%%%%%%%%%%
\begin{eqnarray}
g= \frac{(b-M)(b^2-1)}{b \ln b^2} \equiv g(b,M) \ \ .
\label{g_bM} 
\end{eqnarray}
%%%%%%%%%%%%%%%%%%%%%%%%%%%%%%%%%%%%%%%%%%%%%%%%%%%%%%%%%%%%%%%%%%%%%%%%%
At some specific points, it takes the following values as
%%%%%%%%%%%%%%%%%%%%%%%%%%%%%%%%%%%%%%%%%%%%%%%%%%%%%%%%%%%%
\begin{eqnarray}
\lim_{b \rightarrow 1} g(b,M) = 1-M \ , \ \ 
\lim_{b \rightarrow \infty} g(b,M) = \infty \ \ ,
\nonumber \\
\lim_{b \rightarrow 0} g(b,M) = \left\{
\begin{array}{cl}
0        &(M=0) \\
- \infty &(M \neq 0)
\end{array}
\right. \ \ .
\label{limg_bM} 
\end{eqnarray}
%%%%%%%%%%%%%%%%%%%%%%%%%%%%%%%%%%%%%%%%%%%%%%%%%%%%%%%%%%%%%
Notice that there is a critical difference between asymptotic behaviours for 
$b \rightarrow 0$ in the cases of $M=0$ and $M \neq 0$.  
We draw $g(b,M)$ in Fig. \ref{fig01}.  
%%%%%%%%%%%%%%%%%%%%%%%%%%%%%%%%%%%%%%%%%%%%%%%%%%
%%%%%%%%%%%%%%%%%%%%%%%%%%%%%%%%%%%%%%%%%%%%%%%%%%

\begin{figure}[h]
\centering
\includegraphics[width=10cm]{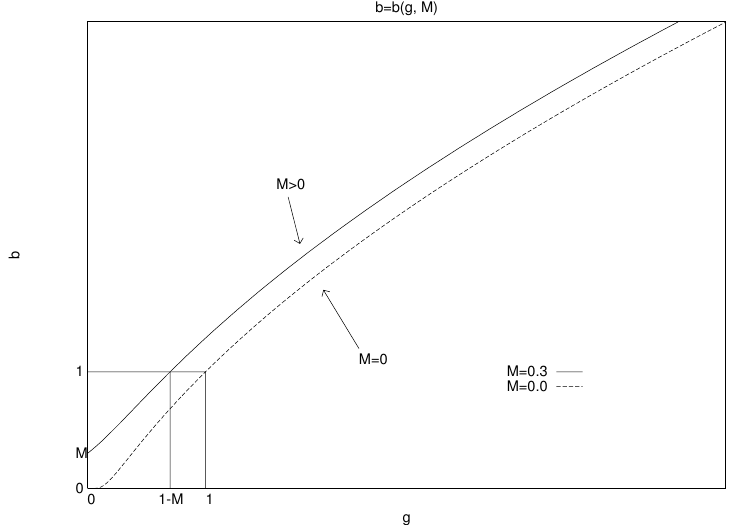}
\caption{ 
The dynamical mass $b=b(g, M)$ for the cases of $M=0$ and $M > 0$.  
This figure is obtained by turning up the figure of $g=g(b, M)$.  
}
\label{fig02}
\end{figure}

Once we know properties of $g(b,M)$, it is easy task to convert them to the 
ones of $b(g,M)$, which is achieved by interchanging the $b$- and $g$-axes 
each other.  
Thus the figure of $b(g,M)$ is obtained by turning up Fig. \ref{fig01}, 
which is presented in Fig. \ref{fig02}.  
This is a key ingredient in getting analytical results without solving 
Eq. (\ref{b}) for $b$.  
%%%%%%%%%%%%%%%%%%%%%%%%%%%%%%%%%%%%%%%%%%%%%%%%%%
%%%%%%%%%%%%%%%%%%%%%%%%%%%%%%%%%%%%%%%%%%%%%%%%%%

From Eq. (\ref{g_bM}), we can show $g(b,M) > g(b,M')$ for a fixed $b$ 
if $M < M'$.  
It means that $b(g,M) < b(g,M')$ for a fixed $g$ if $M < M'$.  
Thus $b(g, M)$ becomes more larger as $M$ increases, when $g$ is fixed.  
There is no intersection between curves for different $M$.  
It concludes that the total mass (the sum of the bare mass and the dynamically 
generated mass) is larger as the bare mass is.  

\subsection{Purely Dynamical Mass} 

\begin{figure}[t]
\centering
\includegraphics[width=10cm]{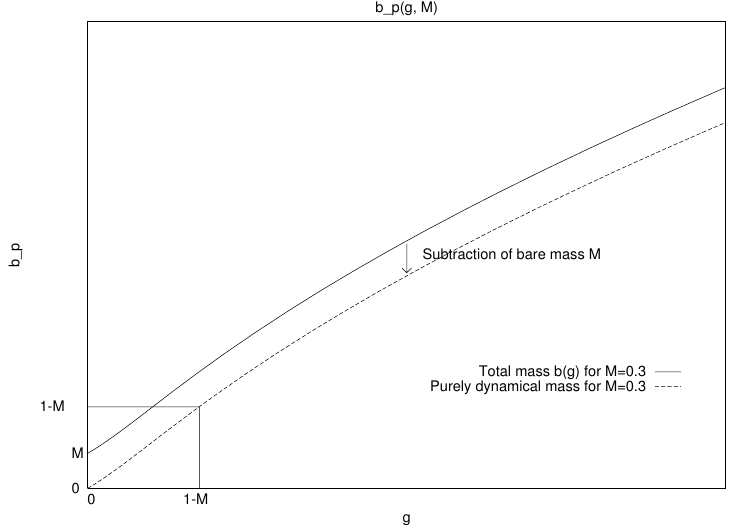}
\caption{ 
The purely dynamical mass $b_p(g)$.  
$b_p$ is defined by the subtraction of the bare mass $M$ from the total 
mass $b$, which corresponds to the parallel transport as shown in the figure.  
}
\label{fig03}
\end{figure}

The $M$-dependence of the total mass is simple as seen above.  
We wonder that there may exist anything new.  
The total mass is the sum of the bare and dynamically generated masses.  
We may consider that the bare mass is an input parameter and the
dynamically generated mass is a response for an external mass insertion.  
It would be interesting to ask how the dynamically generated part depends on 
the bare mass.  
To see that, we define a {\it purely dynamical mass} $B_p$ which is a 
remnant after subtracting the bare mass from the total mass as  
$B_p \equiv B(0) - m$.  
The dimensionless pure dynamical mass $b_p$ is defined by $b_p \equiv 
b(0) - M$.  
This subtraction corresponds to the parallel transport of $g(b,M)$-curve 
to the left in Fig.1 (or $b(g,M)$-curve to the downward in Fig. 2) by $M$.  
Then the relation between $g$ and $b_p$ is given by a function 
$g_p(b_p,M) \equiv g(b_p+M,M)$.  
From Eq. (\ref{g_bM}), we have 
%%%%%%%%%%%%%%%%%%%%%%%%%%%%%%%%%%%%%%%%%%%%%%%%%%%%%%%%%%%%%%%%%%%%%%%%%
\begin{eqnarray}
g = g_p(b_p, M) 
  = \frac{ b_p \{ (b_p + M)^2-1 \} }{ (b_p + M) \ln (b_p + M)^2} \ \ ,
\label{gp_bpM} 
\end{eqnarray}
%%%%%%%%%%%%%%%%%%%%%%%%%%%%%%%%%%%%%%%%%%%%%%%%%%%%%%%%%%%%%%%%%%%%%%%%%
which satisfies the following properties as
%%%%%%%%%%%%%%%%%%%%%%%%%%%%%%%%%%%%%%%%%%%%%%%%%%%%%%%%%%%%%%%%%%%%%%%%%
\begin{eqnarray}
g_p(0,M) = g(M,M)=0 \ , \ \ g_p(b_p,0) = g(b,0) \ \ ,
\nonumber \\ 
g_p(1-M,M) = g_p(1,M) = 1-M \ \ .
\label{limtgpbpM} 
\end{eqnarray}
%%%%%%%%%%%%%%%%%%%%%%%%%%%%%%%%%%%%%%%%%%%%%%%%%%%%%%%%%%%%%%%%%%%%%%%%%
The shape of $g_p(b_p,M)$ is presented in Fig. \ref{fig03}.   
%%%%%%%%%%%%%%%%%%%%%%%%%%%%%%%%%%%%%%%%%%%%%%%%%%
%%%%%%%%%%%%%%%%%%%%%%%%%%%%%%%%%%%%%%%%%%%%%%%%%%

First we study $M$-dependence of $g_p(b_p,M)$ when $g$ is small.  
Since $b_p$ also is expected to be small for $g<<1$, we have 
%%%%%%%%%%%%%%%%%%%%%%%%%%%%%%%%%%%%%%%%%%%%%%%%%%%%%%%%%%%%%%%%%%%%%%%%%%%%%
\begin{eqnarray}
g_p(b_p,M) \cong T(M) b_p \ \ , 
\label{slope} 
\end{eqnarray}
%%%%%%%%%%%%%%%%%%%%%%%%%%%%%%%%%%%%%%%%%%%%%%%%%%%%%%%%%%%%%%%%%%%%%%%%%%%%%
where
%%%%%%%%%%%%%%%%%%%%%%%%%%%%%%%%%%%%%%%%%%%%%%%%%%%%%%%%%%%%%%%%%%%%%%%%%%%%%
\begin{eqnarray}
T(M) \equiv \frac{M^2-1}{2M \ln M} \ \ . 
\label{TM} 
\end{eqnarray}
%%%%%%%%%%%%%%%%%%%%%%%%%%%%%%%%%%%%%%%%%%%%%%%%%%%%%%%%%%%%%%%%%%%%%%%%%%%%%
$T(M)$ presents the $M$-dependence of the slope of $g_p(b_p,M)$ in 
the region of $b_p<<1$.  
At some specific points, $T(M)$ has the following values as 
%%%%%%%%%%%%%%%%%%%%%%%%%%%%%%%%%%%%%%%%%%%%%%%%%%%%%%%%%%%%%%%%%%%%%%%%%%%%%
\begin{eqnarray}
\lim_{M \rightarrow 0} T(M) = \infty \ , \ \ \lim_{M \rightarrow 1} T(M) = 1 \ , \ \
\lim_{M \rightarrow \infty} T(M) = \infty \ \ .  
\label{limTM} 
\end{eqnarray}
%%%%%%%%%%%%%%%%%%%%%%%%%%%%%%%%%%%%%%%%%%%%%%%%%%%%%%%%%%%%%%%%%%%%%%%%%%%%%
It means that $T(M)$ is not a monotonic function of $M$ and may have an 
extrium point.  
To find it, we take the first-derivative of $T(M)$, which is given by 
%%%%%%%%%%%%%%%%%%%%%%%%%%%%%%%%%%%%%%%%%%%%%%%%%%%%%%%%%%%%%%%%%%%%%%%%%%%%%
\begin{eqnarray}
\frac{dT(M)}{dM} &=& \frac{1}{2} \frac{M^2+1}{(M \ln M)^2} S(M^2) \ \ , 
\label{dTdM} 
\end{eqnarray}
%%%%%%%%%%%%%%%%%%%%%%%%%%%%%%%%%%%%%%%%%%%%%%%%%%%%%%%%%%%%%%%%%%%%%%%%%%%%%
where
%%%%%%%%%%%%%%%%%%%%%%%%%%%%%%%%%%%%%%%%%%%%%%%%%%%%%%%%%%%%%%%%%%%%%%%%%%%%%
\begin{eqnarray}
S(x) &\equiv& \frac{1}{2} \ln x - \frac{x-1}{x+1} \ \ . 
\label{SX} 
\end{eqnarray}
%%%%%%%%%%%%%%%%%%%%%%%%%%%%%%%%%%%%%%%%%%%%%%%%%%%%%%%%%%%%%%%%%%%%%%%%%%%%%
Note that the prefactor in front of $S(M^2)$ is positive.  
To see the sign of $S(M^2)$, taking a derivative of $S(x)$, we get 
%%%%%%%%%%%%%%%%%%%%%%%%%%%%%%%%%%%%%%%%%%%%%%%%%%%%%%%%%%%%%%%%%%%%%%%%%%%%%
\begin{eqnarray}
\frac{dS(x)}{dx} = \frac{(x-1)^2}{2x(x+1)^2} \ \ .
\label{dSdx} 
\end{eqnarray}
%%%%%%%%%%%%%%%%%%%%%%%%%%%%%%%%%%%%%%%%%%%%%%%%%%%%%%%%%%%%%%%%%%%%%%%%%
Because $dS(x)/dx$ is positive, $S(x)$ increases monotonically.  
At some specific points, $S(x)$ takes the following values, 
%%%%%%%%%%%%%%%%%%%%%%%%%%%%%%%%%%%%%%%%%%%%%%%%%%%%%%%%%%%%%%%%%%%%%%%%%
\begin{eqnarray}
\lim_{x \rightarrow 0} S(x) = - \infty \ , \ \ S(1) = 0 \ , \ \ 
\lim_{x \rightarrow \infty} S(x) = \infty \ \ .
\label{limSx} 
\end{eqnarray}
%%%%%%%%%%%%%%%%%%%%%%%%%%%%%%%%%%%%%%%%%%%%%%%%%%%%%%%%%%%%%%%%%%%%%%%%%%%%%

\begin{figure}
     \centering
     \includegraphics[width=10cm]{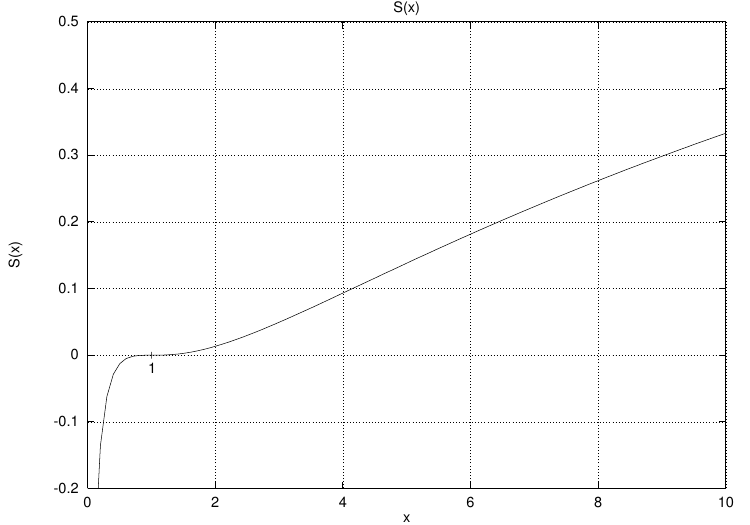}
\caption{ 
The shape of $S(x)$.  
It is a monotonous function having the specific behaviour at $x=1$.  
}
\label{fig04}
\end{figure}

Therefore, we obtain 
%%%%%%%%%%%%%%%%%%%%%%%%%%%%%%%%%%%%%%%%%%%%%%%%%%%%%%%%%%%%%%%%%%%%%%%%%
\begin{eqnarray}
S(x) &\geq& 0 \ \ {\rm when} \ \ x \geq 1 \ \ , 
\nonumber \\ 
S(x) &<& 0 \ \ {\rm when} \ \ 0 < x < 1 \ \ .  
\label{behavSX} 
\end{eqnarray}
%%%%%%%%%%%%%%%%%%%%%%%%%%%%%%%%%%%%%%%%%%%%%%%%%%%%%%%%%%%%%%%%%%%%%%%%%
$S(x)$ is drawn in Fig. \ref{fig04}. 

%%%%%%%%%%%%%%%%%%%%%%%%%%%%%%%%%%%%%%%%%%%%%%%%%%
%%%%%%%%%%%%%%%%%%%%%%%%%%%%%%%%%%%%%%%%%%%%%%%%%%

Combining Eq. (\ref{behavSX}) with Eq. (\ref{dTdM}), we obtain
%%%%%%%%%%%%%%%%%%%%%%%%%%%%%%%%%%%%%%%%%%%%%%%%%%%%%%%%%%%%%%%%%%%%%%
\begin{eqnarray}
\frac{dT(M)}{dM} &>& 0 \ \ {\rm if} \ \ M > 1 \ \ , 
\nonumber \\ 
\frac{dT(M)}{dM} &<& 0 \ \ {\rm if} \ \ M < 1 \ \ ,
\label{behavdTdM} \\
\lim_{M \rightarrow 1} \frac{dT(M)}{dM} &=& 0 \ \ . 
\nonumber
\end{eqnarray}
%%%%%%%%%%%%%%%%%%%%%%%%%%%%%%%%%%%%%%%%%%%%%%%%%%

From Eqs. (\ref{limTM}) and (\ref{behavdTdM}), we know the following behaviour 
of the slope of $g_p(b_p, M)$ in the region of $g<<1$ (or $b_p<<1$):  
When $M=0$, the slope is infinity.  
As $M$ increases, the slope decreases during $0<M<1$, arriving the minimum 
value $\pi/4$ at $M=1$, and turns to increasing if $M>1$.  
These behaviour can be converted to the one of the slope of $b_p(g, M)$ for 
increasing $M$.  
Thus the slope of $b_p(g, M)$ at $M=0$ is equal to zero.  
As $M$ increases, the slope increases until $M=1$ where the slope takes the 
maximum value $\pi/4$.  
Beyond $M=1$, the slope decreases.  
The figure of $b_p(g, M)$ for several M in the region of $g<<1$ (or $b_p<<1$) 
is drawn in Fig. \ref{fig06}. 
The arrows in the figure shows how the dependence of $b_p$ on $g$ is transferred as $M$ increases.

%%%%%%%%%%%%%%%%%%%%%%%%%%%%%%%%%%%%%%%%%%%%%%%%%%
\begin{figure}
     \centering
     \includegraphics[width=10cm]{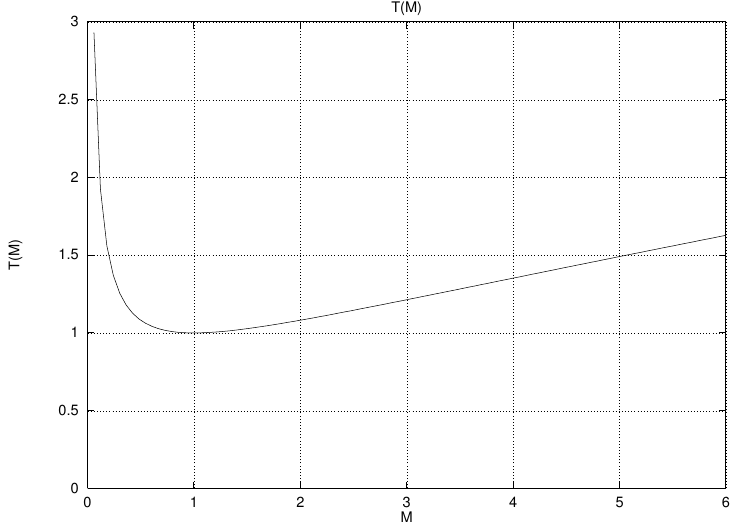}
\caption{ 
The shape of $T(M)$.  
This function has the interesting nature, "duality", that it is invariant 
under $M \rightarrow 1/M$.  
}
\label{fig05}
\end{figure}
%%%%%%%%%%%%%%%%%%%%%%%%%%%%%%%%%%%%%%%%%%%%%%%%%%

%%%%%%%%%%%%%%%%%%%%%%%%%%%%%%%%%%%%%%%%%%%%%%%%%%
\begin{figure}
     \centering
     \includegraphics[width=10cm]{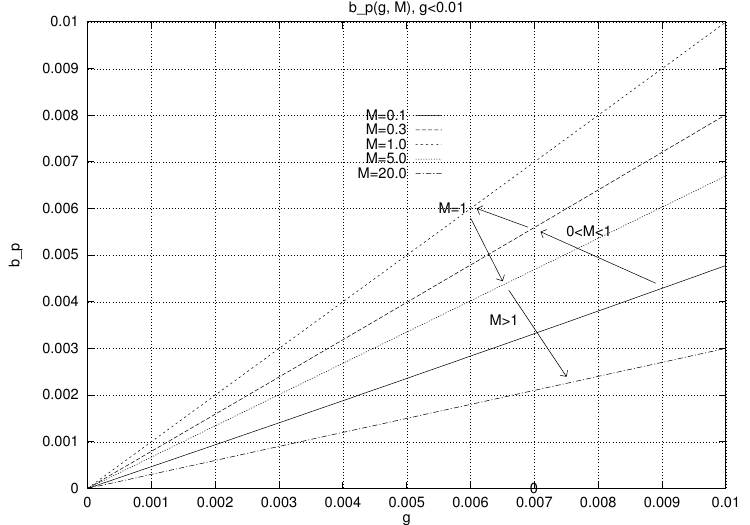}
\caption{ 
$M$-dependence of the slope of $b_p(g, M)$ in the region of $g<<1$ 
under the constant approximation.  
}
\label{fig06}
\end{figure}
%%%%%%%%%%%%%%%%%%%%%%%%%%%%%%%%%%%%%%%%%%%%%%%%%%

The existence of the maximum value of the slope is highly non-trivial.  
It means that there is the maximum value of the purely dynamical mass for a 
fixed $g(<<1)$ when $M$ is changed.  
The purely dynamical mass becomes the maximum at $M=1$ where the fermion 
bare mass is equal to the bare mass of the gauge field.  

In addition, the fact that the slope of $b_p(g, M)$ has the maximum value 
indicates that the slopes for $M(<1)$ coincides with the ones for other 
$M'(>1)$.  
What is the relation between $M$ and $M'$?  
We notice a novel property of the function $T(M)$ as  
%%%%%%%%%%%%%%%%%%%%%%%%%%%%%%%%%%%%%%%%%%%%%%%%%%
\begin{eqnarray}
T(\frac{1}{M}) =  T(M) \ \ .  
\label{dual} 
\end{eqnarray}
%%%%%%%%%%%%%%%%%%%%%%%%%%%%%%%%%%%%%%%%%%%%%%%%%%
It means that the slopes of $b_p(g, M)$ for $M$ and $M'=1/M$ in the region of  
$g(<<1)$ have a same value.  
We may consider that this is a kind of "duality" relation.  
This result is very interesting.  

Now let us turn our attention to a whole region of $g$ or $b$.  
It should be noticed that Eq. (\ref{gp_bpM}) can be expressed by using the 
function $T(M)$ as 
%%%%%%%%%%%%%%%%%%%%%%%%%%%%%%%%%%%%%%%%%%%%%%%%%%
\begin{eqnarray}
g=g_p(b_p,M) = b_p T(b_p+M) \ \ , 
\label{closs} 
\end{eqnarray}
%%%%%%%%%%%%%%%%%%%%%%%%%%%%%%%%%%%%%%%%%%%%%%%%%%
which reduces to Eq. (\ref{slope}) when $b_p<<1$.  
As seen above, the change of the slope of $g_p(b_p, M)$ (or $b_p(g, M)$) in 
the region of $b_p<<1$ (or $g<<1)$) is not nomotonolous when $M$ increases.  
That situation strongly suggests that curves of $g_p(b, M)$ (or $b_p(g, M)$) 
for different $M$ will across each other.  
It would be interesting to seek the purely dynamical mass $b_p$ of satisfying 
$g_p(b_p,M_1)=g_p(b_p,M_2)$, which reduce to $T(b_p+M_1)=T(b_p+M_2)$, for 
$M_1 \neq M_2$.  
At this point, the "duality" relation Eq. (\ref{dual}) plays a key role.  
Thus $g_p(b_p,M_1)=g_p(b_p,M_2)$ is satisfied if 
%%%%%%%%%%%%%%%%%%%%%%%%%%%%%%%%%%%%%%%%%%%%%%%%%%
\begin{eqnarray}
\frac{1}{b_p+M_1} = b_p+M_2 \ \ ,  
\label{cond} 
\end{eqnarray}
%%%%%%%%%%%%%%%%%%%%%%%%%%%%%%%%%%%%%%%%%%%%%%%%%%
and solving this algebraic equation, we have a non-trivial solution, 
%%%%%%%%%%%%%%%%%%%%%%%%%%%%%%%%%%%%%%%%%%%%%%%%%%
\begin{eqnarray}
b_p = \frac{-(M_1+M_2) \pm \sqrt{(M_1-M_2)^2+4}}{2} \ \ .  
\label{solu} 
\end{eqnarray}
%%%%%%%%%%%%%%%%%%%%%%%%%%%%%%%%%%%%%%%%%%%%%%%%%%
If we assume $b>0$, then the condition $M_1 M_2 <1$ should be satisfied.  
Thus we have found that the graphs of $g_p$ for different $M_1$ and $M_2$ 
($M_1 M_2 <1$) across each other at $b_p=\{-(M_1+M_2)+\sqrt{(M_1-M_2)^2+4} \}
/2$.  
Especially if $M_1=M$ and $M_2=0$, then $g_p(b_p, M)$ across $g_p(b_p, 0)$ at 
$b_p=\{ -M+\sqrt{M^2+4} \} /2 (>0)$.  

%%%%%%%%%%%%%%%%%%%%%%%%%%%%%%%%%%%%%%%%%%%%%%%%%%
\begin{figure}
     \centering
     \includegraphics[width=10cm]{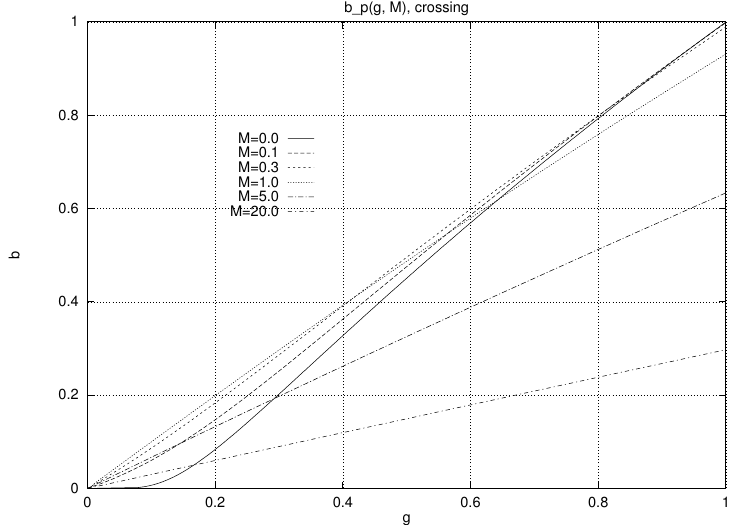}
\caption{ 
Crossing of $b_p(g, M)$ curves for different $M$ 
under the constant approximation.  
}
\label{fig07}
\end{figure}
%%%%%%%%%%%%%%%%%%%%%%%%%%%%%%%%%%%%%%%%%%%%%%%%%%

%%%%%%%%%%%%%%%%%%%%%%%%%%%%%%%%%%%%%%%%%%%%%%%%%%
\begin{figure}
     \centering
     \includegraphics[width=10cm]{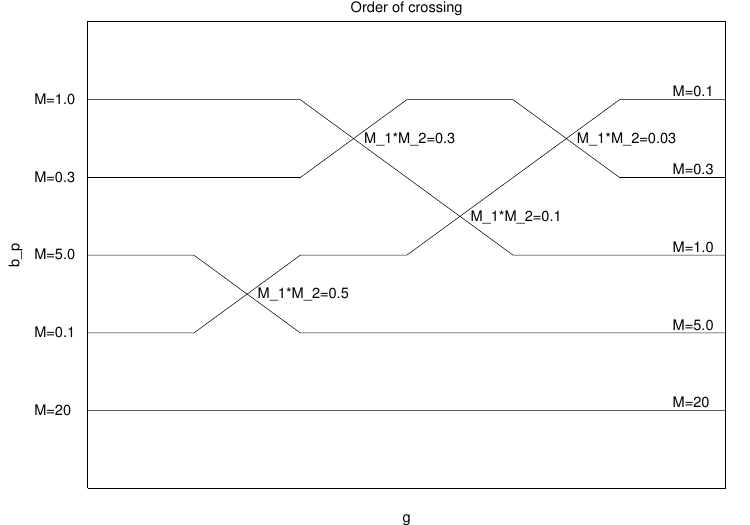}
\caption{ 
Schematic presentation of order of crossings. 
}
\label{fig08}
\end{figure}
%%%%%%%%%%%%%%%%%%%%%%%%%%%%%%%%%%%%%%%%%%%%%%%%%%

%%%%%%%%%%%%%%%%%%%%%%%%%%%%%%%%%%%%%%%%%%%%%%%%%%
\begin{figure}
     \centering
     \includegraphics[width=10cm]{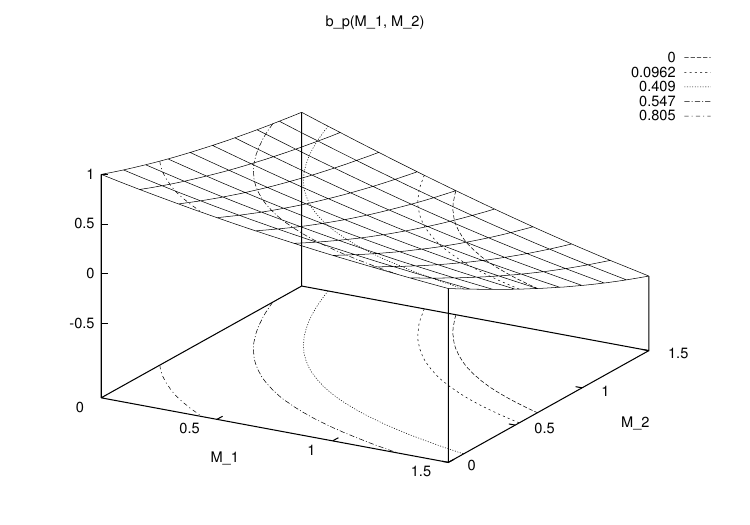}
\caption{ 
($M_1$, $M_2$)-dependence of $b_p$ at crossing points 
under the constant approximation.  
}
\label{fig09}
\end{figure}
%%%%%%%%%%%%%%%%%%%%%%%%%%%%%%%%%%%%%%%%%%%%%%%%%%

\section{Numerical Analysis}
{
We proceed to the more precise numerical analysis.  
We solve the integral equation (\ref{dimless}) directly by using an iterative method [12]. 
As an initial configuration, we input a constant value for all region of the dimensionless 
momentum $t$ as $b(t) = constant$.  
By substituting the constant $b(t)$ into Eq. (\ref{dimless}), we obtain the modified 
function after performing the numerical integration.  
We input again this modified function to the integral equation and get the output function.  
This procedure is repeated until the function converges with a required precision.  
So obtained function is the numerical solution of the integral equation.  
In this section, the results obtained by the analytical method in the constant approximation 
are compared to the ones by the numerical method.

We estimate the purely dynamical mass by the numerical method.  
In the region $0<g<0.01$, we obtain Fig. \ref{fig10}, which is almost same as Fig. \ref{fig06} 
obtained in the constant approximation.

%%%%%%%%%%%%%%%%%%%%%%%%%%%%%%%%%%%%%%%%%%%%%%%%%%
\begin{figure}
     \centering
     \includegraphics[width=10cm]{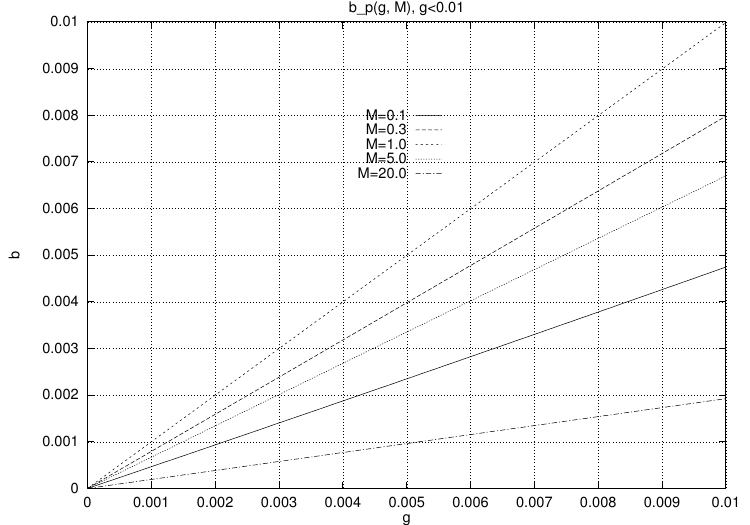}
\caption{ 
The purely dynamical mass in the region of $0 \leq g \leq 0.01$ 
evaluated by the numerical method.  
}
\label{fig10}
\end{figure}
%%%%%%%%%%%%%%%%%%%%%%%%%%%%%%%%%%%%%%%%%%%%%%%%%%

%%%%%%%%%%%%%%%%%%%%%%%%%%%%%%%%%%%%%%%%%%%%%%%%%%%%%%%%%%%%%
\begin{center}
\begin{tabular}{|c|c|c|} 
\hline
Bare mass      &Const. approx. &Numerical             \\ 
($M$)          &($1/T(M)$)     &($b_p/g$ at $g=0.01$) \\ 
\hline
0.1            &0.46517                 &0.47463   \\
0.3            &0.79383                 &0.79810   \\
1.0            &1.00000                 &0.99887   \\
5.0            &0.67060                 &0.67016   \\
20.0           &0.30032                 &0.30029   \\
\hline
\end{tabular}
\end{center}
%%%%%%%%%%%%%%%%%%%%%%%%%%%%%%%%%%%%%%%%%%%%%%%%%%%%%%%%%%%%
\vskip 0.5cm
Table 1: Comparison of slopes $b_p \approx \{ 1/T(M) \} g$ evaluated 
by the analytical method in the constant approximation and by the numerical 
method.  The numerical estimates are smaller than the one obtained in the 
constant approximation by about 36 \%.  

\vskip 1.5cm

%%%%%%%%%%%%%%%%%%%%%%%%%%%%%%%%%%%%%%%%%%%%%%%%%%%%%%%%%%%%%
\begin{center}
\begin{tabular}{|c|c|l|c|l|c|c|} \hline
     &     &        &\multicolumn{2}{c|}{Const. Approx.}
                    &\multicolumn{2}{c|}{Numerical} \\
\cline{4-7}
\multicolumn{1}{|c}{\raisebox{1.5ex}[0pt]{$M_1$}}&
\multicolumn{1}{|c}{\raisebox{1.5ex}[0pt]{$M_2$}}&
\multicolumn{1}{|c|}{\raisebox{1.5ex}[0pt]{$M_1M_2$}}&
$g$     &\multicolumn{1}{|c|}{$b_p$}    &$g$&$b_p$ 
\\ \hline
0.1  &0.3  &0.03  &0.80633 &0.80499 &3.2038 &2.4485 \\
0.1  &1.0  &0.1   &0.56407 &0.54659 &1.7016 &1.4190 \\
0.3  &1.0  &0.3   &0.41757 &0.40948 &1.1412 &1.0034 \\
0.1  &5.0  &0.5   &0.14476 &0.096224&0.22680&0.14988 

\\ \hline
\end{tabular}
\end{center}
%%%%%%%%%%%%%%%%%%%%%%%%%%%%%%%%%%%%%%%%%%%%%%%%%%%%%%%%%%%%%%%
\vskip 0.5cm
Table 2: Comparison of crossing points evaluated by the analytical method 
in the constant approximation and by the numerical method for the case of 
$M_1 M_2 \neq 0$.  

\vskip 1.5cm

To inspect the difference more precisely, we check the values for the slope given by 
Eq. (\ref{slope}) with Eq. (\ref{TM}) noticing $b_p \approx \{ 1/T(M) \}g$.  
The result is given by Table 1 as the comparison of the values between $1/T(M)$ in 
the constant approximation and $b_p/g$ in the numerical method.  
The values have fairly good agreement with the constant approximation.  
This fact suggests that the constant approximation is not so bad to see the tendency of 
the results.  

Concerned with the crossing points which appear for $g<<1$, these points can be found 
for the corresponding points $M_i$ and $M_j$, adopted as an examples in Fig. \ref{fig07}.  
The points are schematically shown more clearly in Fig. \ref{fig08}.  
The comparison of the crossing points evaluated by analytical method in the constant 
approximation and by the numerical method is shown in Table 2.  
Though there are small differences between the values obtained in the constant approximation 
and in the numerical analysis, all values are consistent.  
The analysis confirms the existence of the crossing for the $(g,b_p)$ lines.  

The crossing behavior beyond $g=1$ is shown in Fig. \ref{fig11}, which is derived by the 
numerical analysis.  
For the small $g$, the purely dynamical mass $b_p$ in the case of $M=0$ takes the most 
smallest value and there happens crossing between the values $b_p$'s as $g$ increases.  
Eventually the purely dynamical mass for $M=0$ gets the most largest value.  
We guess that the same thing will happen for a general $g$ if $g$ is large enough, though 
we do not give a proof because the analysis is numerical.  

We compare the crossing points for the purely generated masses with the zero bare mass to 
the ones with the non-zero bare mass by using the constant approximation and numerical method.  
Though the values in constant approximation are smaller than the more precise value given by 
the numerical method, the qualitative behavior of the constant approximation is consistent 
with the numerical results.  

%%%%%%%%%%%%%%%%%%%%%%%%%%%%%%%%%%%%%%%%%%%%%%%%%%
\begin{figure}
     \centering
     \includegraphics[width=10cm]{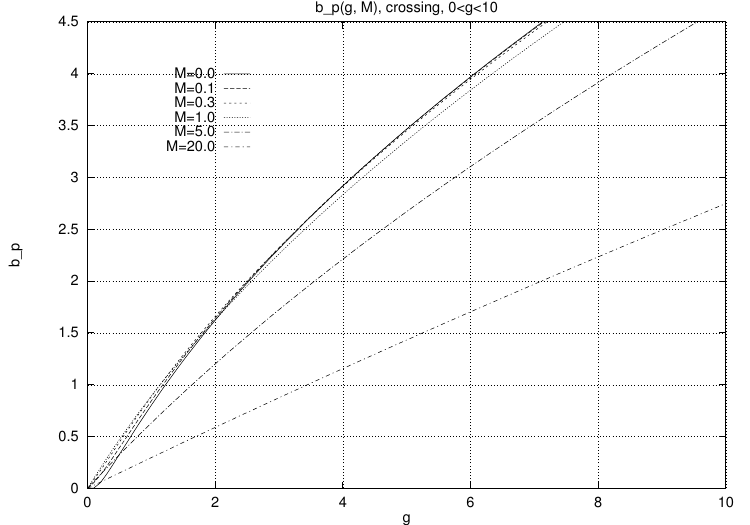}
\caption{ 
The purely dynamical mass in the region of $0 \leq g \leq 10.0$ 
evaluated by the numerical method.  
}
\label{fig11}
\end{figure}
%%%%%%%%%%%%%%%%%%%%%%%%%%%%%%%%%%%%%%%%%%%%%%%%%%

%%%%%%%%%%%%%%%%%%%%%%%%%%%%%%%%%%%%%%%%%%%%%%%%%%%%%%%%%%%%%
\begin{center}
\begin{tabular}{|c|r|c|l|c|c|} \hline
     &              &\multicolumn{2}{c|}{Const. Approx.}
                    &\multicolumn{2}{c|}{Numerical} \\
\cline{3-6}
\multicolumn{1}{|c}{\raisebox{1.5ex}[0pt]{$M_1$}}&
\multicolumn{1}{|c|}{\raisebox{1.5ex}[0pt]{$M_2$}}&
$g$     &\multicolumn{1}{|c|}{$b_p$}    &$g$&$b_p$ 
\\ \hline
0.0  &0.1  &0.95165 &0.95125 &4.3713 &3.1279 \\
0.0  &0.3  &0.86440 &0.86119 &3.6170 &2.6944 \\
0.0  &1.0  &0.56407 &0.54659 &2.0472 &1.6577 \\
0.0  &5.0  &0.29228 &0.19258 &0.48971&0.31843 \\
0.0  &20.0 &0.16635 &0.049876&0.21233&0.063630
\\ \hline
\end{tabular}
\end{center}
%%%%%%%%%%%%%%%%%%%%%%%%%%%%%%%%%%%%%%%%%%%%%%%%%%%%%%%%%%%%%%%
\vskip 0.5cm
Table 3: Comparison of crossing points evaluated by the analytical method 
in the constant approximation and by the numerical method for the case of 
$M_1=0$ and $M_2 \neq 0$.  
\vskip 1.5cm

As a final check, we derive the momentum dependence of the purely dynamical masses $b_p(p)$ 
for various bare masses $M$'s as shown in Fig. \ref{fig12} and Fig. \ref{fig13} by the numerical method.  
The momentum dependent mass is named as a running mass which is a function of momentum.  
The usual dynamical mass is given at the zero momentum.  
We choose the coupling constant $g$ as the corresponding ones which give us the crossing points.  
Then the values of the purely dynamical masses with the corresponding $M$'s agree with the 
suitable values for corresponding $g$.  
The running masses decrease around $p\approx1 \sim 100$.  
Then the running mass function for the larger mass decrease more rapidly than the running mass 
function with the smaller one and the two lines make a shape like a hysteresis curves.  
This behavior is very interesting and suggestive to the duality nature.
}

%%%%%%%%%%%%%%%%%%%%%%%%%%%%%%%%%%%%%%%%%%%%%%%%%%
\begin{figure}
     \centering
     \includegraphics[width=10cm]{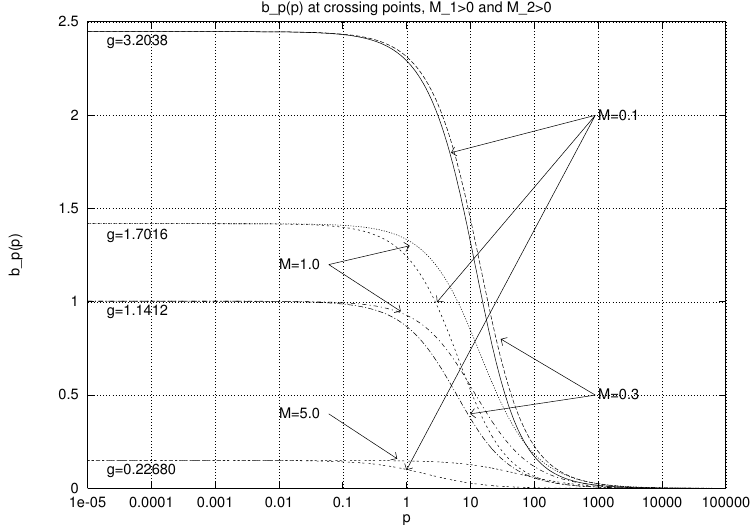}
\caption{
The momentum dependence of the purely dynamical masses $b_p(p)$ for bare masses $0 < M < 20$. See Table 2 for details.
}
\label{fig12}
\end{figure}
%%%%%%%%%%%%%%%%%%%%%%%%%%%%%%%%%%%%%%%%%%%%%%%%%%

%%%%%%%%%%%%%%%%%%%%%%%%%%%%%%%%%%%%%%%%%%%%%%%%%%
\begin{figure}
     \centering
     \includegraphics[width=10cm]{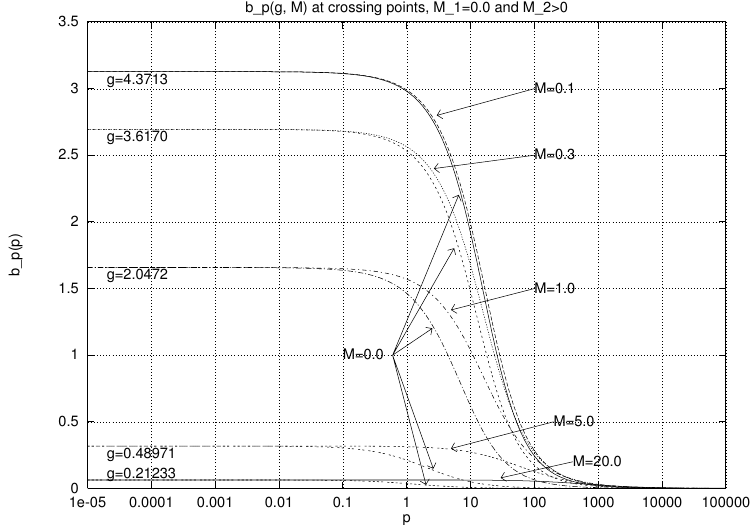}
\caption{
     The momentum dependence of the purely dynamical masses $b_p(p)$ for bare masses $0 < M < 20$. See Table 3 for details.
}
\label{fig13}
\end{figure}
%%%%%%%%%%%%%%%%%%%%%%%%%%%%%%%%%%%%%%%%%%%%%%%%%%

\newpage

\section{Conclusion and discussion}

{
We have studied the mass growth of fermion with the bare mass in (1+1)-dimensional QED in 
which the gauge field also has the bare mass.  
We have used the quenched ladder Schwinger-Dyson equation to estimate the non-perturbative 
effect on the massive fermion coupled to the vector field.  
The effect by the vacuum polarization was not included because of technical difficulties.

We have introduced the purely dynamical mass which is the remaining mass after subtracting 
the bare mass from the total of dynamical mass.  
The mass was newly introduced to see the dynamically induced part in the total mass.

We have used two methods to solve the Schwinger-Dyson equation in the ladder approximation.  
One is a constant approximation in which the momentum variables of the integrand are set to 
zero because the contribution from the zero momentum part is most dominant in the integral 
equation.  
Another method is a numerical analysis by using an iterative method.

For $g<<1$, the purely dynamical mass increases as the coupling does.  
Then we have found the crossing of $(g,b_p)$ lines for the different bare masses.  
Thus the purely dynamical mass is smaller if $g$ is smaller enough.  
But as $g$ increase, $b_p$ with rather smaller bare mass becomes larger.  
There happens the crossing between of $(g,b_p)$ lines and at the crossing point the purely 
dynamical masses for the different bare masses take the same value.  
The reason why this phenomena happen is because of the duality.  
The appearance of the duality is highly nontrivial.

The numerical analysis makes possible for us to know how the crossings happen in all region 
of $g$.  
We obtain the extended version of the duality relation which produces the crossing of 
$(g,b_p)$ lines for the different bare masses.

As a practical problem, it would be interesting to investigate whether this kind of duality 
could be found in realistic condensed matter system.  
A recent nanotechnology in quasi-one-dimensional system could control the strength of an 
effective electric coupling experimentally.  
Then the duality relation between the effective masses could be found.

In future, it would be interesting to include the effects of the vacuum polarization and the 
vertex correction and also to consider the more higher dimensional space-time by extending 
the analysis to the non-abelian cases.  
As the important step, the study in the three-dimensional QED would be interesting [14].
In addition, it would be important to study the 4- or more higher-dimensional theories.  
In higher dimensions, the Schwinger-Dyson technique becomes more problematic in the ultraviolet 
region.  
Therefore it would become important to consider the Higgs mechanism which will be arranged to 
works well in the Schwinger-Dyson method [15].

As a more sophisticated problem it is very challenging to study the vacuum property in the 
mass generation and growth.  
In a relation with the duality, there might be found a changing of any property in the vacuum.  
The duality might be a signal of a phase transition and exists new unknown phases.
}

\section*{Acknowledgment}
The author would like to thank Professor M. Kenmoku for kind hospitality in the Nara 
Women's University where a part of this work has been done.  
He also thank Yuko Fueki for help in a numerical calculation.   
[This is a posthumous paper submitted with the permission of the author’s family.
K. Nakatsugawa, S. Tanda, and G. E. Volovik examined the manuscript and recognized its scientific value, judging it worthy of archival publication as a contribution to physics.
S. Tanda and K. Nakatsugawa further verified its content on behalf of the author to ensure its accuracy.
A. Hashimoto and Yoko Matsuyama recovered the manuscript from the author’s personal computer and assisted in preparing it for submission.]

\newpage

\section*{Appendix A:  Derivation of the Schwinger-Dyson equations}
%%%%%%%%%%%%%%%%%%%%%%%%%%%%%%%%%%%%%%%%%%%%%%%%%%%%%%%%%%%%%%%%%%%%%%%%%%%%%
\begin{eqnarray}
B(p) &=& m+\frac{i}{2} e^2 \int \frac{d^2k}{(2\pi)^2} 
         tr[\gamma^{\mu}S_F(k)\gamma^\nu] D_{\mu \nu}(p-k) \ ,
\nonumber \\
     &=& m- i e^2 \int \frac{d^2k}{(2\pi)^2} \frac{B(k)}{A(k)^{2} k^2 - B(k)^2} 
         \left[ \frac{2}{(p-k)^2-\mu^2} \right.
\nonumber \\
     &&  \left. - \frac{1}{\mu^2} \frac{(p-k)^2}{(p-k)^2 - \mu^2} 
         + \frac{1}{\mu^2} \frac{(p-k)^2}{(p-k)^2 - \lambda \mu^2} \right]
\label{Bd2k} \\
A(p) &=& 1 - \frac{i}{2} e^2 \frac{1}{p^2} \int \frac{d^2k}{(2\pi)^2} 
         tr[\not \! p \gamma^\mu S_F(k) \gamma^\nu] D_{\mu \nu}(p-k)
\nonumber \\
     &=& 1 - i e^2 \frac{1}{p^2} \int \frac{d^2k}{(2\pi)^2} 
         \frac{A(k)}{A(k)^2 k^{2} - B(k)^2} \frac{1}{\mu^2}
\nonumber \\
     & & \left[ \frac{1}{(p-k)^2-\mu^2} 
         - \frac{1}{(p-k)^2- \lambda \mu^2} \right] 
           \{p{\cdot}k (p^2+k^2) - 2 k^2 p^2\}
\label{Ad2k}
\end{eqnarray}
%%%%%%%%%%%%%%%%%%%%%%%%%%%%%%%%%%%%%%%%%%%%%%%%%%%%%%%%%%%%%%%%%%%%%%%%%%%%%
We change the Minkowski metric to the Euclid one by the Wick rotation as
 $k_0 \rightarrow i k_0$ and $\ p_0 \rightarrow i p_0$.  
Then $k^2 = (k_0)^2-(k_1)^2 \rightarrow -(k_0)^2-(k_1)^2 \equiv -k^2$ and 
$p^2 = (p_0)^2-(p_1)^2 \rightarrow -(p_0)^2-(p_1)^2 \equiv -p^2$.  
By the rotation, we get the equations 
%%%%%%%%%%%%%%%%%%%%%%%%%%%%%%%%%%%%%%%%%%%%%%%%%%%%%%%%%%%%%%%%%%%%%%%%%%%%%
\begin{eqnarray}
B(p) &=&m+ e^2 \int \frac{d^2k}{(2\pi)^2} \frac{B(k)}{A(k)^2k^2+B(k)^2}
\nonumber \\
     & & \left[ \frac{2}{(p-k)^2 + \mu^2} 
       + \frac{1}{\mu^2} \frac{(p-k)^2}{(p-k)^2+\mu^2} 
       - \frac{1}{\mu^2} \frac{(p-k)^2}{(p-k)^2+ \lambda \mu^2} \right] \ \ , 
\label{Bw} \\
A(p) &=& 1 - e^2 \frac{1}{p^2} \int \frac{d^2k}{(2\pi)^2} 
             \frac{A(k)}{A(k)^{2}k^2+B(k)^2} \frac{1}{\mu^2} 
\nonumber \\
     & & \left[ \frac{1}{(p-k)^2+\mu^2} 
       - \frac{1}{(p-k)^2+ \lambda \mu^2} \right] 
       \{p{\cdot}k(p^2+k^2)-2k^{2}p^2\} \ \ . 
\label{Aw}
\end{eqnarray}
%%%%%%%%%%%%%%%%%%%%%%%%%%%%%%%%%%%%%%%%%%%%%%%%%%%%%%%%%%%%%%%%%%%%%%%%%%%%%

Here we change the integral variables from Cartesian coordinates to the polar ones.   
%%%%%%%%%%%%%%%%%%%%%%%%%%%%%%%%%%%%%%%%%%%%%%%%%%%%%%%%%%%%%%%%%%%%%%%%%%%%%
\begin{eqnarray}
B(p) &=& m+\frac{e^2}{4\pi^2} \int^\infty_{0} dk 
         \frac{kB(k)}{A(k)^{2}k^2+B(k)^2} 
         \{I_1(\mu^2)+\lambda I_1(\lambda \mu^2)\} 
\label{Bdk} \\ 
A(p) &=& 1 - \frac{e^2}{4\pi^2} \frac{1}{p^2} \int^\infty_{0} dk 
               \frac{kA(k)}{A(k)^{2}k^2+B(k)^2} \frac{1}{\mu^2}
\nonumber \\ 
     & & \left[ (p^2+k^2) \{I_2(\mu^2)-I_2(\lambda \mu^2)\} 
              - 2k^{2}p^2 \{I_1(\mu^2)-I_1(\lambda \mu^2)\} \right]
\label{Adk}
\end{eqnarray}
%%%%%%%%%%%%%%%%%%%%%%%%%%%%%%%%%%%%%%%%%%%%%%%%%%%%%%%%%%%%%%%%%%%%%%%%%%%%%
where $I_i$'s ($i=1, 2$) are the angular integrals as 
%%%%%%%%%%%%%%%%%%%%%%%%%%%%%%%%%%%%%%%%%%%%%%%%%%%%%%%%%%%%%%%%%%%%%%%%%%%%%
\begin{eqnarray}
I_1(a^2) &=& \int^{2\pi}_{0} d\theta \frac{1}{(p-k)^2+a^2} 
          = \frac{2\pi}{\sqrt{(p^2+k^2+\mu^2)^2-4p^2k^2}} 
\label{I1} \\ 
I_2(a^2) &=& \int^{2\pi}_{0} d\theta \frac{p{\cdot}k}{(p-k)^2+a^2} 
          =  \pi \left[ \frac{p^2+k^2+a^2}{\sqrt{(p^2+k^2+a^2)^2-4p^{2}k^2}}
             + 1 \right] 
\label{I2} 
\end{eqnarray}
%%%%%%%%%%%%%%%%%%%%%%%%%%%%%%%%%%%%%%%%%%%%%%%%%%%%%%%%%%%%%%%%%%%%%%%%%%%%%

\section*{Appendix B: Upper bound for $B(0)$}

From Eq. (\ref{Fey}), we can derive an upper bound for $B(0)$.  
We notice that a part of the kernel in Eq. (\ref{Fey}) satisfies an inequality relation 
as 
%%%%%%%%%%%%%%%%%%%%%%%%%%%%%%%%%%%%%%%%%%%%%%%%%%%%%%%%%%%%
\begin{eqnarray}
\frac{2kB(k)}{k^2+B(k)^2} \leq1 \ \ ,
\label{ineq1}
\end{eqnarray}
%%%%%%%%%%%%%%%%%%%%%%%%%%%%%%%%%%%%%%%%%%%%%%%%%%%%%%%%%%%%
because of $(k-B(k))^2 \geq 0$.  
Further, the remaining part in the integral also is bounded by the relation, 
%%%%%%%%%%%%%%%%%%%%%%%%%%%%%%%%%%%%%%%%%%%%%%%%%%%%%%%%%%%%
\begin{eqnarray}
\frac{1}{ \sqrt{(p^2+k^2+\mu^2)^2-4p^2 k^2} } 
\leq \frac{1}{2} \{ \frac{1}{(p-k)^2+\mu^2} + \frac{1}{(p+k)^2+\mu^2} \} \ \ .   
\label{ineq2}
\end{eqnarray}
%%%%%%%%%%%%%%%%%%%%%%%%%%%%%%%%%%%%%%%%%%%%%%%%%%%%%%%%%%%%
By using the inequality relations (\ref{ineq1}) and (\ref{ineq2}), we have 
%%%%%%%%%%%%%%%%%%%%%%%%%%%%%%%%%%%%%%%%%%%%%%%%%%%%%%%%%%%%
\begin{eqnarray}
B(p)&\leq&m+\frac{e^2}{2\pi} \int^\infty_{0} dk \frac{1}{2}
                     \{ \frac{1}{ (p-k)^2+\mu^2 } + \frac{1}{ (p+k)^2+\mu^2 } \} \\ 
\nonumber 
        &=&m+\frac{e^2}{2\pi} \frac{1}{2} \int^\infty_{-\infty} dk 
                   \frac{1}{ (p+k)^2+\mu^2 }  \ \ ,
\nonumber
\end{eqnarray}
%%%%%%%%%%%%%%%%%%%%%%%%%%%%%%%%%%%%%%%%%%%%%%%%%%%%%%%%%%%%
where the integral is easily evaluated as 
%%%%%%%%%%%%%%%%%%%%%%%%%%%%%%%%%%%%%%%%%%%%%%%%%%%%%%%%%%%%
\begin{eqnarray}
\int^\infty_{-\infty} dk \frac{1}{ (p+k)^2+\mu^2 }  = \frac{\pi}{\mu} \ \ ,
\nonumber
\end{eqnarray}
%%%%%%%%%%%%%%%%%%%%%%%%%%%%%%%%%%%%%%%%%%%%%%%%%%%%%%%%%%%%
and finally we have the upper bound for $B(p)$, 
%%%%%%%%%%%%%%%%%%%%%%%%%%%%%%%%%%%%%%%%%%%%%%%%%%%%%%%%%%%%
\begin{eqnarray}
B(p) \leq m+\frac{1}{4} \frac{e^2}{\mu} \ \ ,
\end{eqnarray}
%%%%%%%%%%%%%%%%%%%%%%%%%%%%%%%%%%%%%%%%%%%%%%%%%%%%%%%%%%%%
or, in dimensionless form, 
%%%%%%%%%%%%%%%%%%%%%%%%%%%%%%%%%%%%%%%%%%%%%%%%%%%%%%%%%%%%
\begin{eqnarray}
b(p) \leq M+\frac{\pi}{2} g \ \ . 
\end{eqnarray}
%%%%%%%%%%%%%%%%%%%%%%%%%%%%%%%%%%%%%%%%%%%%%%%%%%%%%%%%%%%%

\section*{Appendix C:  Constant Approximation in arbitrary gauge}

The kernel 
$p^2<<1$ expansion
%%%%%%%%%%%%%%%%%%%%%%%%%%%%%%%%%%%%%%%%%%%%%%%%%%%%%%%%%%%%
\begin{eqnarray}
\frac{1}{\sqrt{(p^2+k^2+\mu^2)^2-4p^2 k^2}} = \frac{1}{k^2+\mu^2} 
+ \frac{k^2-\mu^2}{(k^2+\mu^2)^3} p^2 + O(p^4) 
\label{smallp}
\end{eqnarray}
%%%%%%%%%%%%%%%%%%%%%%%%%%%%%%%%%%%%%%%%%%%%%%%%%%%%%%%%%%%%

Schwinger-Dyson equation at $p=0$
%%%%%%%%%%%%%%%%%%%%%%%%%%%%%%%%%%%%%%%%%%%%%%%%%%%%%%%%%%%%
\begin{eqnarray}
B(0) &=& m+\frac{e^2}{2\pi} \int^\infty_{0} dk \frac{kB(k)}{A(k)^2 k^2+B(k)^2} 
           \left[ \frac{1}{k^2+\mu^2} 
                + \lambda \frac{1}{k^2+ \lambda \mu^2} \right]
\label{B0} \\
A(0) &=& 1 - \frac{e^2}{2\pi} \mu^2 \int^\infty_{0} dk 
             \frac{kA(k)}{A(k)^2 k^2+B(k)^2} 
             \left[ \frac{1}{(k^2+\mu^2)^2} 
           - {\lambda}^2 \frac{1}{(k^2+ \lambda \mu^2)^2} \right]
\label{A0}
\end{eqnarray} 
%%%%%%%%%%%%%%%%%%%%%%%%%%%%%%%%%%%%%%%%%%%%%%%%%%%%%%%%%%%%

Constant approximation: $A(k) \simeq A(0)$, $B(k) \simeq B(0)$ 

Then
%%%%%%%%%%%%%%%%%%%%%%%%%%%%%%%%%%%%%%%%%%%%%%%%%%%%%%%%%%%%
\begin{eqnarray}
B(0) &=& m + \frac{e^2}{2\pi} B(0) 
             \left[ J_1(\mu^2) + \lambda J_1(\lambda \mu^2) \right]
\label{B0J} \\
A(0) &=& 1 - \frac{e^2}{2\pi} \mu^2 A(0) \left[ J_2(\mu^2) 
           - {\lambda}^2 J_2(\lambda \mu^2) \right]
\label{A0J}
\end{eqnarray} 
%%%%%%%%%%%%%%%%%%%%%%%%%%%%%%%%%%%%%%%%%%%%%%%%%%%%%%%%%%%%
where 
%%%%%%%%%%%%%%%%%%%%%%%%%%%%%%%%%%%%%%%%%%%%%%%%%%%%%%%%%%%%
\begin{eqnarray}
J_1(a^2) &=& \int^\infty_0 dk \frac{k}{A(0)^2 k^2 + B(0)^2} \frac{1}{k^2+a^2} \\
&=& \left\{
\begin{array}{ll}
\frac{1}{2} \frac{1}{A(0)^2 a^2 - B(0)^2}  
\ln \left( \frac{A(0)^2 a^2}{B(0)^2} \right) \ \ , & (a^2A^2 \neq B^2) \\
\frac{1}{2B^2} \ \ , & (a^2A^2 = B^2)
\end{array}
\right.
\label{J1} \\
J_2(a^2) &=& \int^\infty_0 dk \frac{k}{A(0)^2 k^2 + B(0)^2}
             \frac{1}{(k^2+a^2)^2} \\
\nonumber \\
&=& \left\{ 
\begin{array}{ll}
\frac{1}{2} \frac{A(0)^2}{(A(0)^2 a^2 - B(0)^2)^2}  
\ln \left( \frac{A(0)^2 a^2}{B(0)^2} \right) 
-  \frac{1}{2 a^2} \frac{1}{A(0)^2 a^2 - B(0)^2} \ \ , & (a^2 A^2 \neq B^2) \\
\frac{1}{4a^2B^2} \ \ , & (a^2 A^2 = B^2)
\end{array}
\right.
\label{J2}
\end{eqnarray} 
%%%%%%%%%%%%%%%%%%%%%%%%%%%%%%%%%%%%%%%%%%%%%%%%%%%%%%%%%%%%

We obtain 
%%%%%%%%%%%%%%%%%%%%%%%%%%%%%%%%%%%%%%%%%%%%%%%%%%%%%%%%%%%%
\begin{eqnarray}
B(0) &=& m + \frac{e^2}{4\pi} B(0) 
        \left[ \frac{1}{A(0)^2 \mu^2 - B(0)^2} 
               \ln \left( \frac{A(0)^2 \mu^2}{B(0)^2} \right) 
        \right. 
\nonumber \\
     & & \left. 
             + \frac{\lambda}{A(0)^2 \lambda \mu^2 - B(0)^2 } 
               \ln \left( \frac{A(0)^2 \lambda \mu^2}{B(0)^2 } \right)
        \right] \ \ ,
\label{B0eq} \\
A(0) &=& 1 - \frac{e^2}{4\pi} \mu^2 A(0)
        \left[ \frac{A(0)^2}{(A(0)^2 \mu^2 - B(0)^2)^2} 
               \ln \left( \frac{A(0)^2 \mu^2}{B(0)^2} \right) 
             - \frac{1}{\mu^2} \frac{1}{A(0)^2 \mu^2 - B(0)^2} 
        \right. 
\nonumber \\
     & & \left. 
             - \frac{\lambda^2 A(0)^2}{(A(0)^2 \lambda \mu^2 - B(0)^2 )^2} 
               \ln \left( \frac{A(0)^2 \lambda \mu^2}{B(0)^2 } \right) 
             + \frac{1}{\mu^2} \frac{\lambda}{A(0)^2 \lambda \mu^2 - B(0)^2 } 
        \right] \ \ ,
\label{A0eq}
\end{eqnarray} 
%%%%%%%%%%%%%%%%%%%%%%%%%%%%%%%%%%%%%%%%%%%%%%%%%%%%%%%%%%%%

When $\lambda=1$, $A(0)=1$ and 
%%%%%%%%%%%%%%%%%%%%%%%%%%%%%%%%%%%%%%%%%%%%%%%%%%%%%%%%%%%%
\begin{eqnarray}
B(0) &=& m + \frac{e^2}{2\pi} \frac{B(0)}{B(0)^2 - \mu^2} 
             \ln \left( \frac{B(0)^2}{\mu^2} \right) 
\label{constB0}
\end{eqnarray} 
%%%%%%%%%%%%%%%%%%%%%%%%%%%%%%%%%%%%%%%%%%%%%%%%%%%%%%%%%%%%

Dimensionless form ($[e]=[M]$): Define
%%%%%%%%%%%%%%%%%%%%%%%%%%%%%%%%%%%%%%%%%%%%%%%%%%%%%%%%%%%%
\begin{eqnarray}
b \equiv \frac{B(0)}{\mu} \ , \ \ 
g \equiv \frac{e^2}{2 \pi \mu^2} \ , \ \ 
M \equiv \frac{m}{\mu}
\end{eqnarray} 
%%%%%%%%%%%%%%%%%%%%%%%%%%%%%%%%%%%%%%%%%%%%%%%%%%%%%%%%%%%%

Then
%%%%%%%%%%%%%%%%%%%%%%%%%%%%%%%%%%%%%%%%%%%%%%%%%%%%%%%%%%%%
\begin{eqnarray}
b = M + g \frac{b}{b^2-1} \ln b^2 
\end{eqnarray} 
%%%%%%%%%%%%%%%%%%%%%%%%%%%%%%%%%%%%%%%%%%%%%%%%%%%%%%%%%%%%
When $g=0$, $b=M$ is a solution of (\ref{b}).  

{
\section*{References}
$^*$ matsuyat@nara-edu.ac.jp

\begin{enumerate}
\item[{[1]}] F.J. Dyson, Phys. Rev. 75 (1949) 1736.
\item[{[2]}] J. Schwinger, Proc. Nat. Acad. Sc. 37 (1951) 452.
\item[{[3]}] J. Schwinger, Phys. Rev. 128 (1962) 2425.
\item[{[4]}] A.L. Fetter and J.D. Walecka. Quantum Theory of Many-particle System. McGraw-Hill (1971) p. 122.
\item[{[5]}] K. Stam, Journal of Phys. G 9 (1983) L229.
\item[{[6]}] R. Delbourgo and G. Thompson. Journal of Phys. G 8 (1982) L185.
\item[{[7]}] S. Coleman, R. Jackiw and L. Susskind, Ann. Phys. (N.Y.) 93 (1975) 267.
\item[{[8]}] S. Coleman, Ann. Phys. (N.Y.) 101 (1976) 239.
\item[{[9]}] J. Lowenstein and A. Swieca, Ann. Phys. (N.Y.) 68 (1971) 172.
\item[{[10]}] T. Maskawa and H. Nakajima, Prog. Theor. Phys. 52 (1974) 1326; 54 (1975) 869.
\item[{[11]}] R. Fukuda and T. Kugo, Nucl. Phys. B 117 (1976) 250.
\item[{[12]}] Y. Hoshino and T. Matsuyama, Phys. Lett. B 222 (1989) 493.
\item[{[13]}] C.D. Fosco and T. Matsuyama, Phys. Lett. B 328 (1994) 513.
\item[{[14]}] P. Maris, Phys. Rev. D54 (1996) 4049.
\item[{[15]}] Ashok Das, Lectures on Quantum Field Theory, p.584 (World Scientific, 2008).
\end{enumerate}
}

\end{document}